\documentclass[showpacs,twocolumn,aps]{revtex4}
\usepackage{graphicx, epsfig}

\newcommand{\be}{\begin{equation}}
\newcommand{\ee}{\end{equation}}
\newcommand{\bea}{\begin{eqnarray}}
\newcommand{\eea}{\end{eqnarray}}

\begin{document}

\title{WaveFunction of the Universe on the Landscape}

\author{Laura Mersini-Houghton}

\affiliation{Department of Physics and Astronomy, UNC-Chapel Hill, NC 27599-3255, USA}

\date{\today}

\begin{abstract} 
This talk reviews the proposal for dynamically selecting the most probable wavefunction of the universe propagating on the landscape of string theory, by means of quantum cosmology. Talk given at 'Albert Einstein Century International Conference'-UNESCO in Paris, July 18-22, 2005.
\end{abstract}


\maketitle

In recent years there has been a radical shift in our understanding of the richness of string theory. Pioneering work done in \cite{vacua} led to the discovery of a multitude of (stable and metastable) vacua solutions with positive or negative energies and compactified $6-$manifolds, coined the landscape by L.Susskind\cite{landscape}. 

Together with the excitement of unraveling a deeper and richer structure of string theory, came a fundamental and puzzling questions: Which vacua in this vastness is 'home' to our universe? One of the most important aspects of any theory is to make predictions that can be tested. The puzzle of finding a selection criterion for our universe which would predict one vacuum among something  like ($10^{100} - 10^{500}$) solutions, currently believed to make up the landscape, is a very challenging task. However addressing this outstanding question offers predictions that can test the theory.

The anthropic selection of some subspace of the landscape, based on environments friendly to life, has received much attention in current literature\cite{weinberg,susskind2,susskind1,lee}.I do not have much to comment about this approach because: it is not clear to me how the probability distribution based on criterion of friendly environments to life can be calculated when, we have an even poorer understanding of life itself; I also do not understand how modern ideas and theories in biology and the science of life, for example the 'fitness landscape' advocated by S.Kauffman\cite{kauffman}, can be accomodated in a consistent picture within the string theory landscape, if conditioned to anthropic probability distributions.

In this talk I focus instead, on an alternative proposal first suggested in \cite{laura1,laura2}, which is based on dynamic selection of the most probable vacua solution on the landscape superspace. The proposal of \cite{laura1,laura2} suggests that we allow the wavefunction of the universe to propagate on the landscape background,(see also \cite{soup,mcinnes,richlaura1,richlaura2} for related approaches), and study this system as a quantum N-body problem. The new selection criterion thus is to predict the most probable wavefunction of the universe based on the structure and dynamics of the string superspace and the dynamics of the wavefunction of the universe propagating through it. I describe the details and some of the results of this approach below.
\subsection{Proposal: Wavefunction of the Universe on the Landscape Background} 

The array of landscape vacua solutions is considered to be a 'lattice'. These vacua are parametrized by a collective coordinate $\phi$ with potential $V(\phi)$. The moduli field $\phi$ collectively characterizes all the internal degrees of freedom in each vacua, but it takes distinct values from one vacuum to another, thereby labelling the vacua 'sites'. Promoting $\phi$ to a collective coordinate can be achieved by taking its value in the $i-th$ vacuum to be the mean center value of the gaussian distributed internal degrees of freedom of that vacuum. These closely spaced resonances are contained within some gaussian width assumed to be very small. The vacua energies of the sites are given by $V(\phi_i)=\lambda_i$.

The proposal of \cite{laura1,laura2} consists on allowing the wavefunction of the universe $\Psi$ to propagate on this landscape background. A probability distribution can be calculated from the solutions of the Wheeler-DeWitt equation (WDW), for the wavefunction of the universe and, the most probable universe be predicted from its peak. Placing the wavefunction of the universe on a real physical background, such as the landscape, is natural since string theory is the leading candidate for describing the realm of quantum gravity. This proposal requires two ingredients: the quantum cosmology framework and, an understanding and knowledge of the landscape structure and distribution of vacua. Quantum cosmology may not yet be an 'airtight' theory and some issues in these formalism are still being debated. Details of the ongoing debate and subtleties can be found in \cite{qcreview}. It is however a perfectly reasonable approach to take, once we assume that quantum mechanics is valid in the ultrahigh-energy regime. On the other hand an understanding of the structure and distribution of vacua in the landcape is currently under intense investigation and much of it remains to be discovered. I comment on the recent progress in this field at the end.

The goal of the proposal is to place quantum cosmology on the landscape background in order to dynamically select the most probable wavefunction of the universe on this background.

Let us define the minisuperspace to be the superspace restricted to the landscape 'lattice' parametrized by the collective moduli coordinate $\phi$ and, to homogenuous and flat 3-geometries with scale factor $a(t)$

\begin{equation}
ds^2= \left[-{\cal N}dt^2+a^2(t)d\bf{x}^2\right],
\label{1}
\end{equation}  
with ${\cal N}$ the lapse function set to one.

The minisuperspace spanned by $(a,\phi)$ is the configuration space in which the wavefunction of the universe $\Psi[a,\phi]$ propagates. 

%%%%%%%%%%%%%%%%%%%%%%%%%%%%%%%%555
The Lagrangian 
for the system with variables $[a,\phi]$ receiving contributions from both: gravity ($L_g$) and, moduli 'superlattice' ($L_{\phi}$), is $L=L_g+L_{\phi}$ where
\begin{eqnarray}
L_g=-\frac{3M_p^2}{8\pi}\frac{a\dot{a}^2}{{\cal N}} \nonumber \\
L_{\phi}=\frac{a^3{\cal N}}{2}\left(-\frac{\dot{\phi}^2}{{\cal N}}-V(\phi)\right)\nonumber\\
\end{eqnarray} 

$\Psi[a,\phi]$ denotes the wavefunctional of the universe propagating on the minisuperspace background $[a,\phi]$.
 The Hamiltonian constraint on the wavefunctional is obtained by varying the combined action with respect to the lapse function $\cal N$, (set to 1 at the end). In the usual manner by promoting $p_a$, $p_{\phi}$ into operators, 
\cite{wdw,Hartle:1983ai,Vachaspati:1988as} the hamiltonian constraint gives the Wheeler-De Witt (WDW) equation \cite{qcreview,wdw},
\begin{eqnarray}
{\hat {\cal H}}\Psi(a,\phi) = 0 ~{\rm with} \nonumber \\
\hat{{\cal H}}=\frac{1}{2e^{3\alpha}}\left[\frac{4\pi}{3M_p^2}
\frac{\partial^2}{\partial\alpha^2}-
\frac{\partial^2}{\partial\phi^2}+e^{6\alpha}V(\phi)\right] \nonumber \\
\label{2}
\end{eqnarray}
where $a$ is replaced by $a=e^{\alpha}$.

$\Psi[\alpha , \phi]$ can be decomposed in modes. By rescaling $\phi$ to $x=e^{3\alpha}\phi$, (and the other relevant quantities in the potential) in order to formally separate variables in Eqn.(\ref{2}) in this simple example we can write, 
\begin{equation}
\Psi({\alpha, x})=\Sigma _j \psi_j(x)F_j(\alpha).
\label{3}
\end{equation}
Replacing (\ref{3}) into (\ref{2}) and using 

\begin{eqnarray}
\hat{\cal{H}}(x)\psi_{j}(x) = \hat \epsilon_j \psi_{j}(x) ~{\rm where} \nonumber  \\
\hat{\cal{H}}(x)=\frac{3M_p^2}{4\pi} \left[
\frac{\partial^2}{\partial x^2}- V(x)\right] \
\nonumber \\
\label{41}
\end{eqnarray}

results in 
\begin{equation}
-\frac{\partial^2}{\partial \alpha ^2}\psi_j(x)F_j(\alpha) = -\hat{\epsilon_j}\psi_jF_j.
\label{4}
\end{equation}
where 'hat' denotes the rescaled $\hat{\epsilon_{j}} = e^{6\alpha} \epsilon_j$ and from hereon$\frac{3M_p^{2}}{4\pi}$ is absorbed into $\epsilon_j$ given in fundamental units.
%%%%%
The $\alpha$ equation of motion is obtained by varying the action $S[\alpha,\phi]$ resulting from Eqn. (\ref{2}), with respect to $\alpha$
\begin{equation}
\ddot{\alpha}+\frac{3}{2}\left[\dot{\alpha}^2+(\dot{x}^2-V(x))e^{-6\alpha}\right]=0 .
\label{alpha}
\end{equation}   
A consistency check shows that the $\alpha$ equation of motion is indeed the Friedman equation for the expansion of the universe born out of the wavefunction $\Psi[\alpha,\phi]$ solutions to WDW equation.

%%%%%%%%
 The 'eigenvalues' $\hat{\epsilon_j}$ are obtained by solving the 'Schrodinger' type equation, Eqn.(~\ref{41}), for 
the field $\psi_{j}(x)$ propagating on the superlattice $V(x)$ with $N$ lattice sites 
${x_i}$ and vacua energies $\lambda_i$.

%%%%%%%%%%%%%%%%%%%%%%%%%%%%%%%%%%
We solve this system as an N-body problem for the multiple scattering of $\Psi[\alpha,x]$ among the N-vacua sites of the landscape. This is done by using the Wigner-Dyson Random Matrix Theory (RMT) methods. The information about the landscape should provide $V(x)$, details of which are unfortunately not known at present. In order to make progress at this stage, a modelling of $V(x)$ is required such that it captures the main features we expect for the $SUSY$ and non$-SUSY$ sectors of the landscape.
 Below I describe the results obtained by applying this proposal to the $SUSY$ and non$-SUSY$ sectors of the landscape and comment briefly on further applications of this formalism. This proposal can be carried out in a straightforward manner once a thorough knowledge of $V(x)$ is derived from string theory.

\subsection{Anderson Localization of the Wavefunction of the Universe on the non-$SUSY$ Sector:} 

 The vacua are located at $\phi_i$, $i=1,...,N$.
 The {\it crucial assumption} made in modelling $V(x)$ is that the vacua energies on the non-$SUSY$ sector are stochastic, namely $\lambda_i$'s are drawn randomly in the interval $\left[0, \pm W \right]$.The energies of the non-$SUSY$ vacua, $\lambda(\phi_i)=\lambda_i$, can take any value in the range $\lambda_i \in \left[0, \pm W \right]$ where $W =O(M_p^{4})$ is an energy  scale related to the Planck or string constant scale. The spacing between different vacua could also be a random 'sprinkling' of sites. In such a setup, the non-$SUSY$ sector of vacua is a randomly disordered lattice of configuration (super-)space of moduli $\phi$, which we name the 'superlattice'.
 The hamiltonian $H(x) = H_{0} + H_{I}$ , with $\phi$ rescaled to $x$, contains two pieces:
the diagonal part $H_{0} = \frac{\partial^2}{\partial x^2} -V_{0}$  where $V_{0}(x_i)=\lambda_i$ and the short range interaction between neighbor vacua [$x_i,x_j$], (the nondiagonal terms), $H_I = V_{I}$.Therefore the potential $V(x)$ includes two terms $V(x)=V_0+V_{I}(x)$.  The short range interaction allows spreading of the wavefunction. Let us assume the nearest neighbor approximation for the short range interaction, for example think of tuneling to the nearest neighbors for simplicity, and introduce a (dimensionless) $\delta$-correlated white noise for the interaction term, i.e $< V_I >=0$ and $< V_{I}(x_i) V_{I}(x_j)> =\Gamma \delta(x_i - x_j)$ where $<...>$ denotes ensemble averaging with respect to [$x_j$].We do not assume $<V_0>=\bar\Lambda$ to be centered around zero.Due to stochasticity, the potential and eigenvalues can not be written in an exact form. 

Following the proposal of\cite{laura1}, the non-$SUSY$ minisuperspace is restricted to the 'disordered superlattice' of vacua $\phi$ for the stochastic non-$SUSY$ sector $V(\phi)$ defined above and, to homogenous flat $3-$geometries with scale factors given by $a(t)$. This is a complicated $N-$body problem because of the multiscattering of $\psi_j(x)$ among many sites of the 'superlattice' but relevant quantities are calculated from probability distributions which can be found exactly.  
Solutions for stochastic backgrounds, Eqn.(\ref{41}), can be found in many papers,\cite{anderson,review1,review2,review3,efemetov}.These investigations show that independent of the specifics of the model, the wavefunctions with energies below a certain scale set by the disorder strength $\Gamma$ do not propagate 
on random lattices with short range interactions. Instead, the wavefunction soon gets localized around a lattice site $x_j$ with 
vacuum energy $\lambda_j$, a phenomenon known as Anderson localization\cite{anderson}.Localization is purely a quantum mechanical effect and it occurs because of the destructive interference of the phases of the wavefunction from multiple scattering among sites. Many examples of this phenomenon have been succesfully applied to lattice QCD, observed experimentally in condensed matter systems and later derived from the (RMT) Wigner-Dyson theory of random matrices\cite{review1,efemetov,review2,review3}, with the $N X N$ matrices obtained over many realizations of the random potential $V(x)$.

The most probable wavefunction of the universe in such stochastic background of the non-SUSY minispuerspace is calculated from the WDW equation, with RMT methods, as follows:
Let us consider the distribution function of the vacua energies from the interval $[0,\pm W]$ by 
$P(\lambda)$. $P(\lambda)$, therefore $P[\hat{H}(x)]$ can be a Gaussian distribution $P(\lambda)=\frac{1}{\sqrt{N\Gamma}}
e^{-\frac{(\lambda-\bar{\Lambda})^2}{N\Gamma}}$, with ensemble averaged mean values $<1|\Lambda_i|N>
=\bar{\Lambda}$ and width $\Gamma =<1|\Gamma_{i}|N>$, or; when disorder is large $\Gamma =O(W)$  
by a flat distribution $P(\lambda)=\frac{1}{2W}$. 
Corrections to the unperturbed energy $\lambda_j$ for the wavefunction localized around $x_j$, along with the evaluated Green's function $<j|G|j> =<j|( \epsilon - \hat{H}_{x} )^{-1}|j>$ can be estimated by the usual perturbation theory \cite{anderson} for weak disorder and by Wigner-Dyson RMT methods for the case when disorder is large and perturbation theory breaks down. The latter is a more elegant and transparent method of calculation since averaging $<...>$ is done by integrating with respect to $\hat{H}(x)$ with probability weight $P[\hat{H}(x)]\simeq e^{ -\frac{{\hat{H}(x)}^{2}}{N\Gamma} }$ \cite{efemetov} rather than ensemble averaging over vacua $[x_j]$, (see reviews\cite{efemetov,review1,review2,review3,anderson} for details).

The eigenfunction $\psi_j(x)$ localized around $x_j$ is given by 
\begin{equation}
|\psi_j(x)|^2\sim \frac{1}{l_j}e^{-\frac{x-x_j}{l_j}}
\label{loc}
\end{equation}
Due to Anderson localization $\psi_{j}(x)$ can not propagate from the non-$SUSY$ sector to other sectors of the landscape. Introducing disorder on the periodic potential of the lattice destroys the constructive interference among the phases, which results in the phenomenon of localization.
The averaged localization length of the system is obtained from the exponential decay of the retarded Green's function and given by the 
ensemble average of the norm of the retarded Green's function $G_R^{-1}$, ($\gamma =<\gamma_j>$), by $l=<l_j>$, $(\frac{l}{L})^2 = \frac{1}{\pi}<1|ln||G^{-1}(x_i,x_j)||N> \simeq (\frac{1}{\gamma})\simeq (\frac{2W}{\Gamma})$ where localization lengths $l_j$ are related to $\gamma_j$
by $l_j\sim \frac{1}{\sqrt{\gamma_j}}$ and $L$ is the size of the landscape sector, $L\simeq Nl_p$.

Replacing the solution Eqn.\ref{41},and Eqn.\ref{energy}, (where for simplicity $\delta_j$ is taken zero), back to WDW Eqn.(\ref{4}), we get $F_j(\alpha)\sim e^{\pm
\sqrt{\hat{\epsilon_j}}\alpha}$ where $\hat{\epsilon}_j$ is the renormalized energy of the $j-th$ state.The wavefunction of the universe solution is localized in the moduli 'superlattice' around some vacua $x_j$ but it has an oscillatory behaviour with respect to $\alpha$.

\begin{equation}
\Psi_{j}(x,\alpha)\simeq \frac{1}{\hat{\epsilon_j}^{1/4} \sqrt{l_j}}e^{\pm i\sqrt{\hat{\epsilon_j}} 
\alpha-\frac{(x-x_j)}{2 l_j}}
\label{5}
\end{equation}  

 The solution for $F_{j}(\alpha)$ in Eqn.(\ref{4})
 is obtained \cite{laura2} by using the Vilenkin boundary condition, with only outgoing modes at future infinity \cite{Vilenkin:1994rn}.Determining the time parameter from the action in the usual manner \cite{wdw} results in 

\begin{equation}
\sqrt{\hat{\epsilon_j}}\alpha = {H_j}t
\label{hubble}
\end{equation}
where $H_j$ is the expansion rate experienced by the local observers bound to the universe $\Psi_j$.

\paragraph*{\bf The Most Probable Universe on the non-$SUSY$ Sector.}

The question: Which solution of Eqn.(\ref{2}) is the most
probable one, can now be addressed statistically by maximizing the density of states (DoS) $\rho(\epsilon_i)$.The single-particle averaged density of states can be obtained from the imaginary part of the advanced Green's function, $Im G_A  (\epsilon_j)\simeq \frac{\gamma_j}{(\epsilon-\epsilon_j)^2 +\gamma_{j}^2}$ with poles at $|\epsilon|=|\lambda_j -i\gamma_j|$ through the expression $\rho(\epsilon)=\frac{1}{\pi}<1|Im G_A|N>$  or more explicitly in RMT from $\rho(\epsilon)= \frac{1}{N}<Tr\delta(\epsilon-H(x))>_{H_{x}} = \frac{1}{N\pi} \int{D(\hat{H_{x}}) P(\hat{H_{x}})Im (G_{A})}$. For our simple 1-dimensional 'superlattice' this equation yields  

\begin{equation}
\rho(\hat{\epsilon})\approx \frac{1}{|\hat{\epsilon}| +\frac{1}{l^2}}
\label{6}
\end{equation}

\begin{figure}[t]
\raggedleft
\centerline{
\epsfxsize=2.5in
\epsfbox{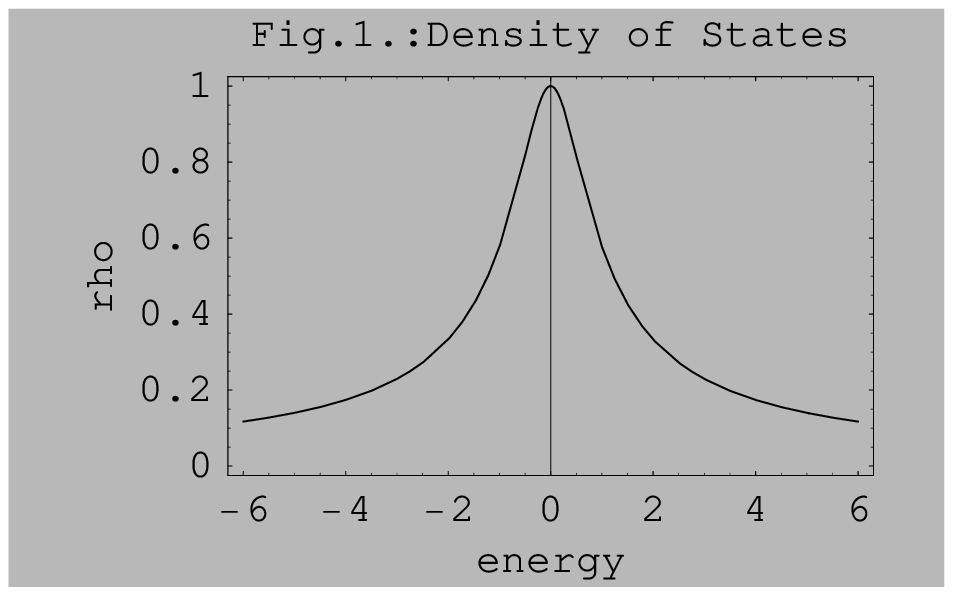}}

\label{fig:1}
\end{figure}

The {\it most probable solution for the wavefunction of the universe} $\Psi_0$ can now be found from the maximum of  
the density of states since $\rho(\epsilon)$ provides the distribution probability of states. From Fig.1, it can be seen that $\rho(\epsilon)$ is peaked around $\epsilon_j\approx 0$ which leads to the conclusion that, with our approach, {\it the most probable universe for the non-$SUSY$ sector}, {\it is the one with a 'physical' cosmological constant $\Lambda_{eff}=\hat{\epsilon}\simeq 0$}. Care should be taken in terms of observers and observables. There are two 'observers' in this framework: a {\it 'local observer'} bound to the wavefunction of the universe $\Psi_{j}[a,\epsilon_j]$ who, (while oblivious about the landscape vacua energies [$\lambda_i$)]), measures $\hat{\epsilon}_j$ as his/her physical cosmological constant in his/her universe because it is only $\epsilon_j$ that determines the expansion rate ${H_j}$ in $F_j(\alpha)$ from $\sqrt{\hat{\epsilon_j}}\alpha = H t$; and a {\it 'superobserver'} bound to the landscape superlattice who can observe all the vacua of the landscape and 'notice' that the wavefunction $\Psi_{j}[a,\epsilon_{j}\simeq 0]$ is localized around some landscape vacuum with some large vacuum energy $\lambda_j$.

DoS is the most useful quantity for extracting statistical predictions about the landscape here: The density of state is a maximum and does not diverge at $\epsilon =0$ due to the breaking of the 'particle-hole' symmetry by gravity.This is the 'celebrated' symmetry breaking of time reversal by the 'intrinsic time' $a(t)$ in the WDW equation.  (If this symmetry were preserved the divergence of $\rho(\hat{\epsilon})$ around $\hat{\epsilon}\simeq 0$ would signal delocalization of $\psi(0)$ thus no universe solution at zero energies); DoS falls off as power-law rather than exponential towards the tail end and has a width $\Gamma$,  thus the probability weight of the higher energy solutions may be non-negligible.

%%%%%%%%%%%%%%%%%%%%%%%%%%%%5
%%%%%%%%%%%%%%%%%%%%%%%%%%%%%%%%%%%%

\subsection{Minisuperspace confined to the SUSY sector of the Landscape:}

The SUSY sector with $N$ vacua sites, is taken to be a regular periodic lattice with vacua sitting at zero energies and spacing of sites of order the string scale $O(l_s)\simeq \frac{1}{\sqrt{\alpha'}}$, \cite{laura1}. The fixed end-point boundary condition was assumed for this sector, which requires that the $s-th$ site satisfy $\Psi_{N+s} = \Psi_s$.

 This 'lattice' configuration contains $N-1$ normal modes. Boundary conditions require that 
\begin{equation}
k_s = \frac{\pi s}{b N}, \ s=1,2,..N.
\label{s}
\end{equation}

Due to the mixing between nearest neighbors from tunneling, the hamiltonian has non-diagonal terms. Diagonalizing the hamiltonian yields the energy eigenvalues of Eqn. (\ref{energy}), thereby splitting the levels and removing the N-fold degeneracy of the ground state. The eigenfunctions obtained after the diagonalization of the hamiltonian give the normal modes of the system $\Psi_{k(s)}(\phi) \simeq sin(k_{s} \phi)$ or, $cos(k_{s} \phi)$. Physically these eigenfunctions are a superposition of left and right moving Bloch plane waves which, due to the constructive intereference in their phases, satisfy the Bragg reflection condition and form standing waves in the minisuperspace lattice of size $L= b N$. The expressions for the standing waves consistent with the boundary condition are 
\begin{equation}
\Psi_s \simeq \frac{sin(k_s \phi)}{\sqrt{k_s}}
\label{standing}
\end{equation}
where the quantum numbers $s$, (not to be confused with lattice site numering), take values in the range  $s=1,..N $. The eigenvalues of the hamiltonian form bands of energy with discrete energy levels, $\epsilon_s$.  A rough estimate for the tunneling rate can be given by $\delta \simeq (\frac{\pi}{b})^2$, known as the mass gap of periodic lattices. The energy of each level with wavenumber $k_s$ is
\begin{equation}
\epsilon_s = 2\delta - 2\delta cos(k_{s} b).
\label{energy}
\end{equation}

The energy eigenvalues $\epsilon_{s}\simeq \frac{\hbar ^2 k_{s}^2}{2}$, and the solutions for the eigenfunctions $\psi_{k_s}$ are given by Eqns. (\ref{s}, \ref{standing}).The lowest energy standing wave is the one for $s=1$, $\tilde{k}_1 = \frac{\pi}{\tilde{b} N}$, $\epsilon_1 = (\frac{\pi}{\tilde{b} N})^2$. By plugging Eqn.(13) back into Eqn. (6) we obtain that $F_{k_s}(\alpha)$ of Eqn.(\ref{6}) has the following solutions
\begin{equation}
F_s(\alpha)\approx \frac{1}{(|\tilde{\epsilon}_s|)^{1/4}}e^{\pm i\sqrt{|\tilde{\epsilon}_s|}}\alpha
\label{10}
\end{equation} 
where the quantized wavenumber $k_s = \frac{\pi s}{bN}$.

The solution to the equation of motion for $\alpha$, Eqn. (12) yields $\alpha = \pm|\epsilon_s|^{1/2}t = \pm (H_s t)$. The growing mode soon dominates over the decaying one, thus we take only the outgoing mode $\alpha = +H_s t$ as our boundary condition at future infinity  (see \cite{Vilenkin:1994rn} for details).Each standing wave mode labelled by the quantum number $s$ in the expression (\ref{6}) for the wavefunction $\Psi(\alpha, \phi)$ describes a DeSitter universe with its own constant nonzero cosmological constant $\tilde{\epsilon_s} \simeq (\frac{\pi s}{b N})^2$, time and expansion rate $\alpha = + H_s t$. The $N-1$ discrete normal modes that form the discrete energy band of bound states, all lifted from zero have respective level energies $\epsilon_s$.The finite gap between energy levels spontaneously breaks the $SUSY$ of the background landscape. Decoherence between levels is resolved since the energy levels are discrete and separated by a finite amount of energy.

The probability ${\cal P}=|\Psi|^2$, up to an overall normalization constant,gives
\begin{equation}
{\cal P} \approx \frac{1}{|\tilde{\epsilon_{s}}|}
\label{p}
\end{equation}
which is peaked around $\Psi_1$ with energy $\epsilon_1 \approx (\frac{\pi}{bN})^2$. Although the SUSY landscape vacua have energies $\lambda=0$, the lifting of the degeneracy of the N-vacua by the wavefunction and thus the spontaneous breaking of SUSY  by the bound state $\Psi_1$ with energy $\epsilon_1$, gives birth to a Universe with a small effective cosmological constant $\Lambda_{eff} =H_1^{2} =\epsilon_1$. $N$ is expected to be large enough. Thus having $\epsilon_1 \approx \Lambda_{eff}$ in the favoured range of $\Lambda_{eff} \approx 10^{-120}M_p^4$ can be easily achieved. Removing the R-symmetry consideration from the $SUSY$ sector of the landscape  would extend this sector to allow the $AdS$ type vacua with $\lambda \leq 0$. However, including the $AdS$ vacua in the superspace does not change the result for the most probable wavefunction of the universe. The $SUSY$ $AdS$ solutions result in a term $\lambda <0$ in Eqn.(13) which may render $F_s(\alpha)$ (\ref{10}) to be a decaying solution for all $\epsilon_s$ for which $\epsilon_s+\lambda<0$. Such solutions have vanishing probability and do not give birth to a Universe. This shows that the most probable solution still remains the first bound state lifted {\it above zero}. The 'standing wave' on the SUSY sector of the landscape, $\Psi_1$, is a unique, stable and most probable solution with nonzero energy. It can therefore be a candidate for the wave function of universe from the landscape.

 It should be noted that when this proposal is applied to the SUSY sector it results in a counterintuitive departure from the point of view taken in literature, where the question ``which vacua do we live in'' implies a highly localized solution for $\Psi_s$. As shown here, the solutions for the SUSY sector are extended solutions over the whole sector, while the solutions for the non-SUSY sector are localized. Therefore, solutions for the wavefunction of the universe do {\it depend on the details of the landscape bacground} in which it propagates. For the SUSY sector while the background itself is supersymetric the wavefunction being lifted from zero, spontaneously breaks SUSY. Local observers bound to $\Psi$ will therefore find $\Lambda \ne 0$ in their universe, while 'superobservers' bound to the landscape superspace find a perfectly supersymetric world.

 \subsection{Discussion}

The discovery of a multitude of vacua solutions with compactified 6-manifolds, $a.k.a$ the landscape of string theory, has had a profound impact on our way of thinking and approach for the cosmological implications of string theory. Initially this vastness of possible universes seemed to imply some disturbing consequences to the falsifiability of the theory itself, unless a selection criterion for picking a unique solution out of the rich landscape, was provided.

The emergence of the landscape presents us with two fundamental aspects that deserve careful investigation:
 
1) Is the multitude of possible universe solutions from string theory or from any theory of quantum gravity, neccessarily a bad thing for that theory? ;

2) Should we postulate a selection principle or derive a selection criterion for choosing our universe among so many possible solutions?

For the first aspect, I would like to argue that the unraveling of the rich structure of the landscape has been good news for string theory and to be expected of any candidate for quantum gravity. The discovered multitude of solutions may indicate that we are on the right track because, fundamental issues in theoretical physics such as: Initial Conditions (IC) for the universe; the origin of constants of nature, quantum numbers and mass scales of our theories; vacuum energy; decoherence and emergence of classicality, an arrow of time and, the observed cosmic coincidences in the late universe \cite{lauraale}, {\it can not}, in fact, be addressed if all we have from quantum gravity is one available sample - our visible universe.{\it The issue, why our universe has picked the parameters it does, unavoidably leads to the question: As compared to what?} Therefore it is almost certain that an underlying theory, embedding our low energy theory of gravity, should contain a more complex space of parameters, with our initial patch emerging from one point of that space. 
It is on these grounds that we expect that, a deeper investigation of cosmological theories from string theory result in a landscape picture. Its discovery is supplementing us for the first time with a real physical background for the phase space of initial conditions, derived from our currently leading theory of quantum gravity, string theory. The emerging landscape may replace our original fuzzy notions of some abstract metauniverse, unknown multiverse,or an abstract phase space of initial conditions, refered to in the past for describing the very early universe. 

The landscape can provide the phase space of initial conditions, as discussed in \cite{richlaura2,richlaura3}, because every vacua solution on the landscape is a potential starting point for a universe with its own parameters, vacuum energy and matter content. Then, how did our patch emerge from the landscape phase space? This question leads us to the second aspect namely, the selection criterion for our universe from the multitude of vacua. In the hope of making any predictions, we have to address why we ended up with our universe, out of so many possible cosmologies. Both attempts, postulating an anthropic selection principle or deriving a dynamic selection mechanism, have been made in the literature. We are still at a very preliminary level of understanding the underlying picture, therefore it may be early to anticipate whether the selection criterion for our universe will be postulated or derived by our physical theories.
%%%%%%%%%%%%%%%%%%%%%%%%%%%%%%%55
The purpose of the proposal discussed here is to offer {\it a selection criterion for the landscape vacua which is derived from the dynamics of the wavefunction of the universe propagating on the landscape background}. The selection mechanism was derived by placing quantum cosmology on the landscape background, thereby calculating the most probable wavefunction of the universe from solutions to the WDW equation\cite{qcreview,wdw}. Initially this selection rule was applied to the $SUSY$ sector of the landscape, (reviewed in Sec.C.), and the most probable universes were 'standing wave' solutions extended over the whole landscape $SUSY$ sector, peaked around energies $\Lambda\simeq \frac{1}{N^2}$, with $N$ refering to the number of vacua\cite{laura1}.

The same proposal applied to the investigation of the minisuperspace restricted to the non-$SUSY$ sector of the landscape and flat 3-geometries was reviewed in Sec.C. In the absence of knowledge about the detailed structure of the non-$SUSY$ sector, vacua energies were considered to be a {\it stochastic} variable, namely randomly drawn from the interval $[0,\pm W]$. Wavefunction solutions found from WDW equation, Eqn.\ref{6}, exhibit the well-known phenomenon of Anderson localization characteristic of disordered systems. Localization of the wavepacket ensures that coherence of the wavefunction of the universe is maintained over large time-scales. In the non-SUSY sector of the landscape our findings indicate that the most probable wavefunction of the universe solution selects 
states of zero energy and consequently of zero 'physical' cosmological constant, (although the superlatice 'bare' vacua energies $\lambda_i$ where the most probable wavefunction localizes, may be as large as the disorder strength $W$, $\lambda_j\approx O(\gamma)$).

Based on the important results for the vacua distribution by Denef et al.\cite{douglas}, an improved and a more realistic model for the landscape structure which included the internal vacua degrees of freedom was later given in \cite{richlaura2,richlaura3}. Solutions on this more complex background were found by exploring the analogy with condensed matter systems of the same universality class \cite{altland}. This approach was then used to investigate and make predictions about the initial conditions and entropy of inflationary solutions from the landscape phase space\cite{richlaura2,richlaura3}.
 Many issues remain to be addressed yet. Future directions for the application of our proposal have to involve an extension of the minisuperspace to more degrees of freedom as well as complementing the search for the string theory signatures\cite{observsign,tde} by connecting the landscape picture to astrophysical observables.

\end{document}